\documentclass[12pt]{article}
\begin{document}

\begin{center}
{\Large\bf{}An unique alternative non-negative gravitational
energy tensor to the Bel-Robinson tensor in the quasilocal small
sphere limit}
\end{center}

\begin{center}
Lau Loi So\\
Department of Physics, National Central University, Chung-Li 320, Taiwan\\
Department of Physics, Tamkang University, Tamsui 251, Taiwan\\
(Dated on 28 June 2010, \quad s0242010@gmail.com)
\end{center}

\begin{abstract}
The Bel-Robinson tensor $B_{\alpha\beta\mu\nu}$ gives a positive
definite gravitational energy in the quasilocal small sphere limit
approximation. However, there is an alternative tensor
$V_{\alpha\beta\mu\nu}$ that was proposed recently that offers the
same positivity as $B_{\alpha\beta\mu\nu}$ does.  We have found
that $V_{\alpha\beta\mu\nu}$ is the unique alternative tensor with
$B_{\alpha\beta\mu\nu}$ which implies that these two tensors are a
basis for expressions that have the desirable non-negative
gravitational energy in the small sphere limit. In other words,
the `energy-momentum' density according to $B_{\alpha\beta\mu\nu}$
and $V_{\alpha\beta\mu\nu}$ are on equal footing at the same
limit.
\end{abstract}

\section{Introduction}
The Bel-Robinson tensor $B_{\alpha\beta\mu\nu}$ has many nice
properties. It is completely symmetric, completely trace free and
completely divergence free.  It is usually regarded as being
related to gravitational energy.  In particular, the gravitational
energy-momentum density in the small sphere vacuum limit is
generally expected to be proportional to the Bel-Robinson tensor.
This expectation is related to the requirement of energy
positivity \cite{Szabados}.

It should be emphasized that, generally speaking, a positive
energy proof for a quasilocal expression is not easy. Here we
consider specifically pseudotensor expressions. In fact, the
quasilocal methods are not fundamentally different from
pseudotensor methods \cite{ChangNesterChen}. The gravitational
energy expression in a small region limit can be investigated
through the pseudotensors. Normally, the expansion of a
pseudotensor expression up to second order can be represented by
certain tensors $B_{\alpha\beta\mu\nu}$, $S_{\alpha\beta\mu\nu}$
and $K_{\alpha\beta\mu\nu}$ \cite{SoCQG2009,MTW}.  In other words,
quasilocal expression is a fancy name for a pseudotensor. Even
though a pseudotensor is not a tensorial object, this does not
imply that it is useless. The second order expansion expression
provides guidance as to whether the gravitational energy
expression is positive or not. More precisely, a negative
quasilocal gravitational energy expression on a small scale
definitely guarantees that it can be negative on the large scale.
Conversely, a positive energy expression in the small region
implies that there may be a chance to obtain positivity in a large
region.

However, it is natural to question whether studying this kind of
quasilocal formulation has physical significance. It is well known
that the gravitational energy density cannot be detected at a
point because of the equivalence principle (see section 20.4 in
\cite{MTW}). Fortunately, the quasilocal idea has physical
meaning, i.e., the gravitational energy density is well defined at
the quasilocal level theoretically
\cite{Hawking,Penrose1982,Horowitz}. Practically, pseudotensors
can be used to calculate the tidal heating \cite{Purdue} (e.g.,
Jupiter and Io) as well as using the quasilocal formalism
\cite{Booth}. Moreover, from the last decade, there are many
researchers who believe that finding a good quasilocal expression
(especially one that is locally positive) is meaningful and
worthwhile, see e.g., \cite{Szabados,LiuYau} and the many
references contained therein.

In the past, the Bel-Robinson tensor has been considered to be the
only tensor which contributes positivity in the small sphere
limit. However, we recently found another tensor \cite{SoCQG2009},
$V_{\alpha\beta\mu\nu}$ (defined in (\ref{24Feb2009}) below),
which is also quadratic in the curvature, and which enjoys the
same positivity properties as $B_{\alpha\beta\mu\nu}$.  More
precisely, the associated 4 momentum in a small sphere is a
Lorentz-covariant future pointing non-spacelike vector (see 4.2.2
of \cite{Szabados}).  For short we call this property casuality.
Furthermore, we found that $B_{\alpha\beta\mu\nu}$ and
$V_{\alpha\beta\mu\nu}$ are a basis for expressions which have the
desirable non-negative gravitational energy in the small sphere
vacuum limit. As we found, $V_{\alpha\beta\mu\nu}$ fulfills the
weak energy condition which gives positive energy in the small
sphere limit, this is then sufficient to argue that the
`energy-momentum' density according to $B_{\alpha\beta\mu\nu}$ and
$V_{\alpha\beta\mu\nu}$ are on an equal footing in the same limit.
We will prove in section 3 that $V_{\alpha\beta\mu\nu}$ is the
unique alternative tensor that has the desirable positive energy
property in the small region limit.

In this work, we examine some properties of
$V_{\alpha\beta\mu\nu}$ and some other quadratic in curvature
tensors, $S_{\alpha\beta\mu\nu}$, $K_{\alpha\beta\mu\nu}$ and
$W_{\alpha\beta\mu\nu}$, which have shown up in the expansion of
energy in the small sphere limit. For instance, it has already
been shown that $V_{\alpha\beta\mu\nu}$ does not have the dominant
energy condition \cite{SoCQG2009}. We also found another tensor
$V'_{\alpha\beta\mu\nu}$, which is not restricted to the
pseudotensor conservation of energy-momentum requirement, but does
satisfy the weak energy condition requirement.

Here we are concerned with finding a suitable form for a
pseudotensor (equivalently, a quasilocal Hamiltonian expression,
see \cite{SoNesterCQG2009}) in the small region limit. For the
zeroth order term, the pseudotensor gives the mass density as the
equivalence principle demands.  Moreover, we also need to consider
the ADM mass at the spatial infinity.  Combining these two
constraints \cite{SoCQG2009} can confine a suitable pseudotensor
expression. Finally, the non-vanishing second order terms
contribute the gravitational energy-momentum in a small region
limit; these terms are quadratic in the curvature tensor.

\section{Quadratic curvature tensors}
There are three basic tensors that commonly occur in the
gravitational pseudotensor expression \cite{MTW,SoIJMPD}
\begin{eqnarray}
B_{\alpha\beta\mu\nu}
&:=&R_{\alpha\lambda\mu\sigma}R_{\beta}{}^{\lambda}{}_{\nu}{}^{\sigma}
+R_{\alpha\lambda\nu\sigma}R_{\beta}{}^{\lambda}{}_{\mu}{}^{\sigma}
-\frac{1}{8}g_{\alpha\beta}g_{\mu\nu}\mathbf{R}^{2},\\
S_{\alpha\beta\mu\nu}
&:=&R_{\alpha\mu\lambda\sigma}R_{\beta\nu}{}^{\lambda\sigma}
+R_{\alpha\nu\lambda\sigma}R_{\beta\mu}{}^{\lambda\sigma}
+\frac{1}{4}g_{\alpha\beta}g_{\mu\nu}\mathbf{R}^{2},\label{29April2009}\\
K_{\alpha\beta\mu\nu}
&:=&R_{\alpha\lambda\beta\sigma}R_{\mu}{}^{\lambda}{}_{\nu}{}^{\sigma}
+R_{\alpha\lambda\beta\sigma}R_{\nu}{}^{\lambda}{}_{\mu}{}^{\sigma}
-\frac{3}{8}g_{\alpha\beta}g_{\mu\nu}\mathbf{R}^{2},
\end{eqnarray}
where $\mathbf{R}^{2}=R_{\rho\tau\xi\kappa}R^{\rho\tau\xi\kappa}$.
Some properties of $S_{\alpha\beta\mu\nu}$ and
$K_{\alpha\beta\mu\nu}$ \cite{SoCQG2009} that are easily verified
using the above definitions and the well known vacuum identity
\cite{Yefremov}
$R_{\alpha\lambda\sigma\tau}R_{\beta}{}^{\lambda\sigma\tau}
=\frac{1}{4}g_{\alpha\beta}\mathbf{R}^{2}$ are
\begin{eqnarray}
&&S_{\alpha\beta\mu\nu}\equiv{}S_{(\alpha\beta)(\mu\nu)}\equiv{}S_{(\mu\nu)(\alpha\beta)},\quad\quad{}
S_{\alpha\beta\mu}{}^{\mu}\equiv{}\frac{3}{2}g_{\alpha\beta}\mathbf{R}^{2}
,\quad\quad~{}S_{\alpha\mu\beta}{}^{\mu}\equiv{}0,\label{9aSep2009}\\
&&K_{\alpha\beta\mu\nu}\equiv{}K_{(\alpha\beta)(\mu\nu)}\equiv{}K_{(\mu\nu)(\alpha\beta)},\quad{}
K_{\alpha\beta\mu}{}^{\mu}\equiv{}-\frac{3}{2}g_{\alpha\beta}\mathbf{R}^{2},
\quad{}K_{\alpha\mu\beta}{}^{\mu}\equiv{}0.\label{9bSep2009}
\end{eqnarray}
Note that unlike $B_{\alpha\beta\mu\nu}$, both
$S_{\alpha\beta\mu\nu}$ and $K_{\alpha\beta\mu\nu}$ are neither
totally symmetric nor totally trace free.  Obviously,
(\ref{9aSep2009}) and (\ref{9bSep2009}) already indicate that
$S_{\alpha\beta\mu\nu}$ and $K_{\alpha\beta\mu\nu}$ do not have
the completely trace free property.  For the non-completely
symmetric property, one can verify this by using Petrov type D
\cite{Petrov}.  In particular, we found a case where
\begin{eqnarray}
S_{0011}\neq{}S_{0101},\quad{} K_{0011}\neq{}K_{0101},
\end{eqnarray}
where we have evaluated them using the electric part $E_{ab}$ and
magnetic part $H_{ab}$, defined in terms of the Weyl tensor
\cite{Carmeli} as follows:
\begin{equation}
E_{ab}:=C_{a0b0},\quad{}H_{ab}:=*C_{a0b0}.
\end{equation}

In order to appreciate the nice properties of the Bel-Robinson
tensor, we compare some components of $S_{\alpha\beta\mu\nu}$ and
$K_{\alpha\beta\mu\nu}$.  For the analog of the electromagnetic
stress tensor energy density,
\begin{eqnarray}
B_{0000}&=&E_{ab}E^{ab}+H_{ab}H^{ab},\\
S_{0000}&=&2(E_{ab}E^{ab}-H_{ab}H^{ab}),\\
K_{0000}&=&-E_{ab}E^{ab}+3H_{ab}H^{ab}.
\end{eqnarray}
Likewise for the momentum density (i.e., the Poynting vector)
\begin{equation}
B_{000i}=2\epsilon_{ijk}E^{jd}H^{k}{}_{d},\quad{}
S_{000i}=0,\quad{}
K_{000i}=2\epsilon_{ijk}E^{jd}H^{k}{}_{d}.
\end{equation}
Finally, the stress,
\begin{eqnarray}
B_{00ij}&=&\delta_{ij}(E_{ab}E^{ab}+H_{ab}H^{ab})-2(E_{id}E_{j}{}^{d}+H_{id}H_{j}{}^{d}),\\
S_{00ij}&=&2\left[\delta_{ij}(-E_{ab}E^{ab}+H_{ab}H^{ab})
+4(-E_{id}E_{j}{}^{d}+H_{id}H_{j}{}^{d})\right],\\
K_{00ij}&=&\delta_{ij}(E_{ab}E^{ab}-3H_{ab}H^{ab})-4E_{id}E_{j}{}^{d}.
\end{eqnarray}
From the above comparison, it is clear that the Bel-Robinson
tensor indeed has the best analog with the electromagnetic stress
tensor $T^{\mu\nu}$.  In detail, in Minkowski coordinates
$(\tau,x,y,z)$ the components of the electrodynamic stress tensor
are
\begin{eqnarray}
T^{00}&=&\frac{1}{2}\left(E_{a}E^{a}+B_{a}B^{a}\right),\\
T^{0i}&=&\delta^{ij}\epsilon_{jab}E^{a}B^{b}~=~(\vec{E}\times\vec{B})^{i},\\
T^{ij}&=&\frac{1}{2}\left[\delta^{ij}\left(E_{a}E^{a}+B_{a}B^{a}\right)
-2\left(E^{i}E^{j}+B^{i}B^{j}\right)\right].
\end{eqnarray}
where $\vec{E}$ and $\vec{B}$ refer to the electric and magnetic
field density.

Using a Taylor series expansion, the metric tensor can be written
as
\begin{equation}
g_{\alpha\beta}(x)
=g_{\alpha\beta}(0)+\partial_{\mu}g_{\alpha\beta}(0)x^{\mu}
+\frac{1}{2}\partial^{2}_{\mu\nu}g_{\alpha\beta}(0)x^{\mu}x^{\nu}+\ldots.
\end{equation}
At the origin in Riemann normal coordinates (RNC)
\begin{eqnarray}
g_{\alpha\beta}(0)&=&\eta_{\alpha\beta},
\quad\quad\quad\quad\quad\quad\quad\quad~~
\partial_{\mu}g_{\alpha\beta}(0)=0, \\
-3\partial^{2}_{\mu\nu}g_{\alpha\beta}(0)
&=&R_{\alpha\mu\beta\nu}+R_{\alpha\nu\beta\mu},\quad\quad
-3\partial_{\nu}\Gamma^{\mu}{}_{\alpha\beta}(0)
=R^{\mu}{}_{\alpha\beta\nu}+R^{\mu}{}_{\beta\alpha\nu}.
\end{eqnarray}
For the quadratic curvature tensors, there are 4 independent basis
\cite{Deser} expressions, we may use
\begin{eqnarray}
\tilde{B}_{\alpha\beta\mu\nu}
&:=&R_{\alpha\lambda\mu\sigma}R_{\beta}{}^{\lambda}{}_{\nu}{}^{\sigma}
+R_{\alpha\lambda\nu\sigma}R_{\beta}{}^{\lambda}{}_{\mu}{}^{\sigma}
=B_{\alpha\beta\mu\nu}+\frac{1}{8}g_{\alpha\beta}g_{\mu\nu}\mathbf{R}^{2},\label{20aJan2009}\\
\tilde{S}_{\alpha\beta\mu\nu}
&:=&R_{\alpha\mu\lambda\sigma}R_{\beta\nu}{}^{\lambda\sigma}
+R_{\alpha\nu\lambda\sigma}R_{\beta\mu}{}^{\lambda\sigma}
=S_{\alpha\beta\mu\nu}-\frac{1}{4}g_{\alpha\beta}g_{\mu\nu}\mathbf{R}^{2},\label{29aApril2009}\\
\tilde{K}_{\alpha\beta\mu\nu}
&:=&R_{\alpha\lambda\beta\sigma}R_{\mu}{}^{\lambda}{}_{\nu}{}^{\sigma}
+R_{\alpha\lambda\beta\sigma}R_{\nu}{}^{\lambda}{}_{\mu}{}^{\sigma}
=K_{\alpha\beta\mu\nu}+\frac{3}{8}g_{\alpha\beta}g_{\mu\nu}\mathbf{R}^{2},\\
\tilde{T}_{\alpha\beta\mu\nu}
&:=&-\frac{1}{8}g_{\alpha\beta}g_{\mu\nu}\mathbf{R}^{2}.\label{20bJan2009}
\end{eqnarray}
These four tensors are manifestly symmetric in the last two
indices which means
$\tilde{M}_{\alpha\beta\mu\nu}=\tilde{M}_{\alpha\beta(\mu\nu)}$.
Then it automatically imply
$\tilde{M}_{\alpha\beta(\mu\nu)}=\tilde{M}_{(\alpha\beta)(\mu\nu)}$.
Moreover, it also naturally turns out that
$\tilde{M}_{(\alpha\beta)(\mu\nu)}=\tilde{M}_{(\mu\nu)(\alpha\beta)}$.
Explicitly, they fulfill the symmetry
$\tilde{M}_{\alpha\beta\mu\nu}\equiv\tilde{M}_{(\alpha\beta)(\mu\nu)}
\equiv\tilde{M}_{(\mu\nu)(\alpha\beta)}$. Although there exists
some other tensors different from $\tilde{B}_{\alpha\beta\mu\nu}$,
$\tilde{S}_{\alpha\beta\mu\nu}$, $\tilde{K}_{\alpha\beta\mu\nu}$
and $\tilde{T}_{\alpha\beta\mu\nu}$, they are just linear
combinations of these four.  For instance
\begin{equation}
\tilde{T}_{\alpha\mu\beta\nu}+\tilde{T}_{\alpha\nu\beta\mu}
\equiv\tilde{B}_{\alpha\beta\mu\nu}
+\frac{1}{2}\tilde{S}_{\alpha\beta\mu\nu}
-\tilde{K}_{\alpha\beta\mu\nu}
+2\tilde{T}_{\alpha\beta\mu\nu}.\label{7Nov2008}
\end{equation}
The above identity can be obtained by making use of the completely
symmetric property of the Bel-Robinson tensor. Using
(\ref{7Nov2008}), we can rewrite the Bel-Robinson tensor in a
different representation \cite{Deser}:
\begin{equation}
B_{\alpha\beta\mu\nu}\equiv-\frac{1}{2}S_{\alpha\beta\mu\nu}
+K_{\alpha\beta\mu\nu}+\frac{5}{8}g_{\alpha\beta}g_{\mu\nu}\mathbf{R}^{2}
-\frac{1}{8}(g_{\alpha\mu}g_{\beta\nu}+g_{\alpha\nu}g_{\beta\mu})\mathbf{R}^{2}.
\label{29Oct2008}
\end{equation}
This equation will be used in section 3.

\section{An unique alternative non-negative gravitational energy tensor in small sphere limit}

\subsection{Proof the unique alternative non-negative energy tensor $V_{\alpha\beta\mu\nu}$}
Using RNC Taylor series expansion around any point (i.e., at any
preselected point we consider a small coordinate sphere in RNC,
see e.g., \cite{Szabados,Horowitz,Bergqvist,Garecki}), consider
all the possible combinations of the small region energy-momentum
density in vacuum.  In the neighbourhood of any preselected point
in RNC, the pseudotensor then has the form \cite{SoCQG}
\begin{equation}
2\kappa\,t_{\alpha}{}^{\beta}=2G_{\alpha}{}^{\beta
}+\left(a_{1}\tilde{B}_{\alpha}{}^{\beta}{}_{\mu\nu}
+a_{2}\tilde{S}_{\alpha}{}^{\beta}{}_{\mu\nu}
+a_{3}\tilde{K}_{\alpha}{}^{\beta}{}_{\mu\nu}
+a_{4}\tilde{T}_{\alpha}{}^{\beta}{}_{\mu\nu}
\right)x^{\mu}x^{\nu}+{\cal{}O}(\mbox{Ricci},x)+{\cal{}O}(x^{3}),
\label{25Sep2008}
\end{equation}
where $\kappa=8\pi{}G/c^{4}$ (here we take units such that $c=1$
for simplicity) and $a_{1}$ to $a_{4}$ are real numbers. Here
$G_{\alpha}{}^{\beta}$ is the Einstein tensor, but we will
consider the vacuum case, so
\begin{equation}
G_{\alpha}{}^{\beta}=\kappa\,T_{\alpha}{}^{\beta}=0.
\end{equation}
Then the first order linear in Ricci terms
${\cal{}O}({\mbox{Ricci}},x)$ vanish.  The lowest order
non-vanishing term is of second order, compared to this in the
small sphere limit we can ignore the third order terms
${\cal{}O}(x^{3})$. From now on, the second order term will be
kept but the others are dropped. The essential purpose of the
present paper is to prove that $B_{\alpha\beta\mu\nu}$ and
$V_{\alpha\beta\mu\nu}$, where \cite{SoCQG2009}
\begin{equation}
V_{\alpha\beta\mu\nu}:=S_{\alpha\beta\mu\nu}+K_{\alpha\beta\mu\nu},\label{24Feb2009}
\end{equation}
are a basis for positive gravitational energy in the small sphere
limit. There are two physical conditions which can constrain the
unlimited combinations between $\tilde{B}_{\alpha\beta\mu\nu}$,
$\tilde{S}_{\alpha\beta\mu\nu}$, $\tilde{K}_{\alpha\beta\mu\nu}$
and $\tilde{T}_{\alpha\beta\mu\nu}$.  The first one is the
conservation of the energy-momentum density and the second is the
positive
gravitational energy in the small sphere limit.\\
$(\mathbf{i})$ First condition: energy-momentum conservation.
Consider (\ref{25Sep2008}) as follows
\begin{eqnarray}
0&=&\partial_{\beta}\,t_{\alpha}{}^{\beta}\nonumber\\
&=&\left(a_{1}\tilde{B}_{\alpha}{}^{\beta}{}_{\mu\nu}
+a_{2}\tilde{S}_{\alpha}{}^{\beta}{}_{\mu\nu}
+a_{3}\tilde{K}_{\alpha}{}^{\beta}{}_{\mu\nu}
+a_{4}\tilde{T}_{\alpha}{}^{\beta}{}_{\mu\nu}\right)(\delta^{\mu}_{\beta}x^{\nu}+x^{\mu}\delta^{\nu}_{\beta})\nonumber\\
&=&2\left(a_{1}\tilde{B}_{\alpha}{}^{\beta}{}_{\mu\beta}
+a_{2}\tilde{S}_{\alpha}{}^{\beta}{}_{\mu\beta}
+a_{3}\tilde{K}_{\alpha}{}^{\beta}{}_{\mu\beta}
+a_{4}\tilde{T}_{\alpha}{}^{\beta}{}_{\mu\beta}\right)x^{\mu}
\nonumber\\
&=&\frac{1}{4}(a_{1}-2a_{2}+3a_{3}-a_{4})g_{\alpha\beta}x^{\beta}\mathbf{R}^{2}.\label{24eFeb2010}
\end{eqnarray}
Therefore, the constraint for the conservation of the
energy-momentum density is
\begin{equation}
a_{4}=a_{1}-2a_{2}+3a_{3}.\label{9Jan2009}
\end{equation}
Although there are an infinite number of combinations which can
fulfill the above constraint, it has removed one degree of
freedom. As each single tensor of $\tilde{B}_{\alpha\beta\mu\nu}$,
$\tilde{S}_{\alpha\beta\mu\nu}$, $\tilde{K}_{\alpha\beta\mu\nu}$
or $\tilde{T}_{\alpha\beta\mu\nu}$ cannot satisfy the conservation
requirement, but a linear combination of them can. One can
simplify the situation by eliminating
$\tilde{T}_{\alpha\beta\mu\nu}$ which is absorbed by
$\tilde{B}_{\alpha\beta\mu\nu}$, $\tilde{S}_{\alpha\beta\mu\nu}$
and $\tilde{K}_{\alpha\beta\mu\nu}$. Then there are only 3 basis
tensors left.  Thus one can rewrite (\ref{25Sep2008}) as
\begin{eqnarray}
2\kappa\,t_{\alpha}{}^{\beta} &=&\left[a_{1}(\tilde{B}_{\alpha}{}^
{\beta}{}_{\mu\nu} +\tilde{T}_{\alpha}{}^ {\beta}{}_{\mu\nu})
+a_{2}(\tilde{S}_{\alpha}{}^ {\beta}{}_{\mu\nu}
-2\tilde{T}_{\alpha}{}^ {\beta}{}_{\mu\nu})
+a_{3}(\tilde{K}_{\alpha}{}^ {\beta}{}_{\mu\nu}
+3\tilde{T}_{\alpha}{}^ {\beta}{}_{\mu\nu})\right]
x^{\mu}x^{\nu}\nonumber\\
&=& \left(a_{1}B_{\alpha}{}^{\beta}{}_{\mu\nu}
+a_{2}S_{\alpha}{}^{\beta}{}_{\mu\nu}
+a_{3}K_{\alpha}{}^{\beta}{}_{\mu\nu}
\right)x^{\mu}x^{\nu}.\label{25aSep2008}
\end{eqnarray}
Paying attention to \cite{SoNesterCQG2009,PRD}, when we consider
all the possible expressions for the pseudotensors which including
the flat metric, there does appear a linear combination of these
three tensors. We defined $K_{\alpha\mu\nu}$ \cite{SoIJMPD} just
for convenience and without any physical reason. In the beginning,
it seems an interesting and even a mysterious combination that why
only $B_{\alpha\beta\mu\nu}$, $S_{\alpha\beta\mu\nu}$ and
$K_{\alpha\beta\mu\nu}$ always showed up in the expression.
Nothing more, nothing less. Now, we discovered that it is not an
accident but becomes a necessary requirement because only
$B_{\alpha\beta\mu\nu}$, $S_{\alpha\beta\mu\nu}$ and
$K_{\alpha\beta\mu\nu}$ can satisfy the condition of the
energy-momentum density conservation.\\
$(\mathbf{ii})$ Second condition: non-negative gravitational
energy in the small sphere limit. The purpose of the pseudotensor
is for determining the gravitational energy-momentum, the
associated energy-momentum can be calculated as
\begin{eqnarray}
2\kappa\,P_{\mu}&=&\int_{t=0}t^{\rho}{}_{\mu\xi\kappa}x^{\xi}x^{\kappa}d\Sigma_{\rho}
~=~t^{0}{}_{\mu{}lm}\int_{t=0}x^{l}x^{m}d^{3}x\nonumber\\
&=&t^{0}{}_{\mu{}lm}\frac{\delta^{lm}}{3}\int{}r^{2}d^{3}x
~=~t^{0}{}_{\mu{}l}{}^{l}\,\frac{4\pi{}r^{5}}{15},\label{5Jan2009}
\end{eqnarray}
where $l,m=1,2,3$. Using this calculation method, the
energy-momentum in the small sphere limit for (\ref{25aSep2008})
becomes
\begin{equation}
P_{\mu}=(-E,\vec{P})=-\frac{r^{5}}{60G}\left(
a_{1}B_{\mu{}0l}{}^{l}+a_{2}S_{\mu{}0l}{}^{l}+a_{3}K_{\mu{}0l}{}^{l}\right).\label{25dSep2008}
\end{equation}
The `energy-momentum' values associated with
$B_{\alpha\beta\mu\nu}$, $S_{\alpha\beta\mu\nu}$ and
$K_{\alpha\beta\mu\nu}$  are proportional to
\begin{eqnarray}
B_{\mu{}0l}{}^{l}&=&(E_{ab}E^{ab}+H_{ab}H^{ab},2\epsilon_{cab}E^{ad}H^{b}{}_{d}),\label{2aMar2010}\\
S_{\mu{}0l}{}^{l}&=&-10(E_{ab}E^{ab}-H_{ab}H^{ab},0),\label{25bSep2008}\\
K_{\mu{}0l}{}^{l}&=&B_{\mu{}0l}{}^{l}-S_{\mu{}0l}{}^{l}.\label{25cSep2008}
\label{25cSep2008}
\end{eqnarray}

Looking back at (\ref{25aSep2008}), we are interested in the
positive gravitational energy within a small sphere limit, the
Bel-Robinson tensor already satisfies this condition. Precisely
\begin{equation}
B_{00l}{}^{l}=E_{ab}E^{ab}+H_{ab}H^{ab}\geq{}0.
\end{equation}
The rest of the job is to find the coefficients $a_{2}$ and
$a_{3}$.  Using (\ref{25cSep2008}), rewrite (\ref{25dSep2008}) as
\begin{equation}
P_{\mu}=-\frac{r^{5}}{60G}\left[
(a_{1}+a_{3})B_{\mu{}0l}{}^{l}+(a_{2}-a_{3})S_{\mu{}0l}{}^{l}\right].\label{27Sep2008}
\end{equation}
Equation (\ref{25bSep2008}) shows that $S_{\mu{}0l}{}^{l}$ cannot
ensure positivity, since we should allow for any magnitude of
$||E_{ab}||$ and $||H_{ab}||$. In other words, for
$S_{\alpha\beta\mu\nu}$ the sign of the `energy' density is
uncertain. Therefore the only possibility for (\ref{27Sep2008}) to
guarantee positivity is when $a_{1}+a_{3}\geq{}10|a_{2}-a_{3}|$.
However, if we consider that the momentum should be future
pointing and non-spacelike (i.e., inside the light cone such that
$-P_{0}\geq|\vec{P}|$), the unique requirement for
(\ref{27Sep2008}) to assure non-negative is when $a_{2}=a_{3}$. In
other words, causality. Moreover, using (\ref{24Feb2009}) and
(\ref{25cSep2008}), we obtained
\begin{equation}
V_{\mu{}0l}{}^{l}=S_{\mu{}0l}{}^{l}+K_{\mu{}0l}{}^{l}=B_{\mu{}0l}{}^{l}.\label{2bMar2010}
\end{equation}
Consequently (\ref{25dSep2008}) becomes
\begin{equation}
P_{\mu}=-\frac{r^{5}}{60G}\left(
a_{1}B_{\mu{}0l}{}^{l}+a_{2}V_{\mu{}0l}{}^{l}\right)
=-\frac{r^{5}}{60G}(a_{1}+a_{2})B_{\mu{}0l}{}^{l}.
\end{equation}
Hence the proof is completed.  Indeed $B_{\alpha\beta\mu\nu}$ and
$V_{\alpha\beta\mu\nu}$ are a basis for expressions which have
non-negative gravitational `energy' density in vacuum.

\subsection{Physical meaning of the completely traceless property of
$B_{\beta\mu\nu}$ and $V_{\alpha\beta\mu\nu}$}

For the quasilocal small sphere region, there are four fundamental
quadratic Weyl curvature tensors.  We know that
$\tilde{M}_{\alpha\beta\mu\nu}=\tilde{M}_{(\alpha\beta)(\mu\nu)}=\tilde{M}_{(\mu\nu)(\alpha\beta)}$
for all
$\tilde{M}_{\alpha\beta\mu\nu}\in\{\tilde{B}_{\alpha\beta\mu\nu},
\tilde{S}_{\alpha\beta\mu\nu},\tilde{K}_{\alpha\beta\mu\nu},\tilde{T}_{\alpha\beta\mu\nu}\}$.
In order to check the completely trace free property of the linear
combinations of $\tilde{B}_{\alpha\beta\mu\nu},
\tilde{S}_{\alpha\beta\mu\nu},\tilde{K}_{\alpha\beta\mu\nu}$ and
$\tilde{T}_{\alpha\beta\mu\nu}$.  We only need to consider two
cases.\\
Case (i).  Consider the trace on the first and third indices:
\begin{equation}
a_{1}\tilde{B}^{\alpha}{}_{\mu\alpha\nu}
+a_{2}\tilde{S}^{\alpha}{}_{\mu\alpha\nu}
+a_{3}\tilde{K}^{\alpha}{}_{\mu\alpha\nu}
+a_{4}\tilde{T}^{\alpha}{}_{\mu\alpha\nu}
=\frac{1}{8}(a_{1}-2a_{2}+3a_{3}-a_{4})g_{\mu\nu}\mathbf{R}^{2}.\label{24bFeb2010}
\end{equation}
Case (ii).  Consider the trace on the first pair:
\begin{equation}
a_{1}\tilde{B}^{\alpha}{}_{\alpha\mu\nu}
+a_{2}\tilde{S}^{\alpha}{}_{\alpha\mu\nu}
+a_{3}\tilde{K}^{\alpha}{}_{\alpha\mu\nu}
+a_{4}\tilde{T}^{\alpha}{}_{\alpha\mu\nu} =\frac{1}{2}(a_{1}+a_{2}
-a_{4})g_{\mu\nu}\mathbf{R}^{2}.\label{24aFeb2010}
\end{equation}
For completely traceless, (\ref{24bFeb2010}) and
(\ref{24aFeb2010}) have to vanish at the same time
\begin{eqnarray}
a_{1}-2a_{2}+3a_{3}-a_{4}&=&0,\label{24dFeb2010}\\
a_{1}+a_{2} -a_{4}&=&0.\label{24cFeb2010}
\end{eqnarray}
Notice that (\ref{24dFeb2010}) is the same constraint for the
energy-momentum conservation expressed in (\ref{9Jan2009}). This
means that one of the mathematical trace free condition turns out
to be one of the physical criteria. Solving the above two
equations, we recovered the same requirement for the gravitational
energy-momentum (i.e., casuality) which was indicated in
(\ref{27Sep2008}), explicitly $a_{2}=a_{3}$. Moreover, using this
totally traceless property at the quasilocal small sphere limit
shown in (\ref{24dFeb2010}) and (\ref{24cFeb2010}), we recovered
the same result: that there are two tensors that generate the
basis mentioned in section 3.1. The proof is follows
\begin{eqnarray}
&&a_{1}\tilde{B}_{\alpha\beta\mu\nu}
+a_{2}\tilde{S}_{\alpha\beta\mu\nu}
+a_{3}\tilde{K}_{\alpha\beta\mu\nu}
+a_{4}\tilde{T}_{\alpha\beta\mu\nu}\nonumber\\
&=&a_{1}(\tilde{B}_{\alpha\beta\mu\nu}+\tilde{T}_{\alpha\mu\nu})
+a_{2}(\tilde{S}_{\alpha\beta\mu\nu}
+\tilde{K}_{\alpha\beta\mu\nu}
+\tilde{T}_{\alpha\beta\mu\nu})\nonumber\\
&=&a_{1}B_{\alpha\beta\mu\nu} +a_{2}V_{\alpha\beta\mu\nu},
\end{eqnarray}
where
\begin{equation}
V_{\alpha\beta\mu\nu}=\tilde{S}_{\alpha\beta\mu\nu}
+\tilde{K}_{\alpha\beta\mu\nu} +\tilde{T}_{\alpha\beta\mu\nu}.
\end{equation}
The completely traceless property turns out to be a remarkable
result because it is not just a mathematical property, but it
reveals some physical meaning and conditions. Namely, the
conservation of the energy-momentum and casuality.

In other words, we have discovered necessary and sufficient
conditions. From considering the quasilocal small sphere limit, we
find the completely traceless for these two fourth rank tensors
$B_{\alpha\beta\mu\nu}$ and $V_{\alpha\beta\mu\nu}$ guarantees the
fulfillment of the energy-momentum conservation and casuality.
Conversely, if any fourth rank tensor satisfies the conservation
of energy-momentum and casuality, then it must be a linear
combination of $B_{\alpha\beta\mu\nu}$ and
$V_{\alpha\beta\mu\nu}$.

\subsection{Counting the independent components of
$B_{\alpha\beta\mu\nu}$ and $V_{\alpha\beta\mu\nu}$} As
$B_{\alpha\beta\mu\nu}$ and $V_{\alpha\beta\mu\nu}$ are different
but share the same gravitational `energy-momentum' density, then
one may interested to know how many non-vanishing independent
components $B_{\alpha\beta\mu\nu}$ and $V_{\alpha\beta\mu\nu}$
have.  Using (\ref{29Oct2008}) and (\ref{24Feb2009}), here we
write the alternative relationship between $B_{\alpha\beta\mu\nu}$
and $V_{\alpha\beta\mu\nu}$ \cite{SoCQG2009}
\begin{equation}
V_{\alpha\beta\mu\nu}:=B_{\alpha\beta\mu\nu}+W_{\alpha\beta\mu\nu},\label{3aMar2010}
\end{equation}
where we have defined
\begin{equation}
W_{\alpha\beta\mu\nu}:=\frac{3}{2}S_{\alpha\beta\mu\nu}
-\frac{5}{8}g_{\alpha\beta}g_{\mu\nu}\mathbf{R}^{2}
+\frac{1}{8}(g_{\alpha\mu}g_{\beta\nu}+g_{\alpha\nu}g_{\beta\mu})\mathbf{R}^{2}.
\end{equation}
Basically, these three tensors ($B$, $V$, $W$) are fourth rank, in
principle they could have 256 components. However, after
considering the symmetry properties, they only have a relatively
small amount of independent components.  It may be important to do
this simple counting because it reduces the workload (i.e.,
computer algebra) when one calculates all the components of these
three tensors.

First of all, we count the number of components of
$B_{\alpha\beta\mu\nu}$.  In principle, as $B_{\alpha\beta\mu\nu}$
is completely symmetric, by explicit examination it reduces to 35
components. There is a formula that directly gives this number. A
$k\,$th rank totally symmetric tensor in $n$ dimensional space has
$C^{n+k-1}_{k}$ components. For our case
\begin{equation}
C^{4+4-1}_{4}=35.
\end{equation}
Since $B_{\alpha\beta\mu\nu}$ is completely trace free, there are
10 constraints which can be replaced by the other components.
Finally, we only left 25 independent components for
$B_{\alpha\beta\mu\nu}$ (see \cite{Petrov}).

Secondly, we count $V_{\alpha\beta\mu\nu}$. Tensor
$V_{\alpha\beta\mu\nu}$ does not have the completely symmetric
property, but fulfills some certain symmetries
$V_{\alpha\beta\mu\nu}=V_{(\alpha\beta)(\mu\nu)}=V_{(\mu\nu)(\alpha\beta)}$.
In principle, this reduces $V_{\alpha\beta\mu\nu}$ to 55
components. However, when we consider the totally trace free
property of $V_{\alpha\beta\mu\nu}$ (but not completely
symmetric), then there are two extra constraints need to be taken
into account
\begin{equation}
V^{\alpha}{}_{\alpha\mu\nu}=0, \quad\quad{}
V^{\alpha}{}_{\mu\alpha\nu}=0.
\end{equation}
Hence we have $55-10-10=35$ independent components for
$V_{\alpha\beta\mu\nu}$.

Finally, for the completeness, we count $W_{\alpha\beta\mu\nu}$.
Note that $W_{\alpha\beta\mu\nu}$ and $V_{\alpha\beta\mu\nu}$ are
similar. In detail
\begin{equation}
W_{\alpha\beta\mu\nu}=W_{(\alpha\beta)(\mu\nu)}=W_{(\mu\nu)(\alpha\beta)},\quad{}
W^{\alpha}{}_{\alpha\mu\nu}=0,\quad{}W^{\alpha}{}_{\mu\alpha\nu}=0,\quad{}
\end{equation}
In principle, there should be at most 35 components. However, one
must consider the extra constraint
\begin{equation}
W_{\alpha\beta\mu\nu}+W_{\alpha\mu\nu\beta}+W_{\alpha\nu\beta\mu}=0.
\end{equation}
Finally we have $35-25=10$ independent components for
$W_{\alpha\beta\mu\nu}$.

\subsection{Physical application for $V_{\alpha\beta\mu\nu}$} The
physical application of $V_{\alpha\beta\mu\nu}$ is similar to
$B_{\alpha\beta\mu\nu}$ in the small region limit.  For instance,
the Einstein pseudotensor does not have a suitable positivity
expressions as the sign of the energy density is uncertain. Recall
the result for Einstein in vacuum \cite{SoNesterCQG2009}
\begin{equation}
2\kappa\,E_{\alpha}{}^{\beta}=\frac{1}{18}
(4B_{\alpha}{}^{\beta}{}_{\mu\nu}-S_{\alpha}{}^{\beta}{}_{\mu\nu})x^{\mu}x^{\nu}.
\end{equation}
Referring to (\ref{5Jan2009}), the corresponding gravitational
energy is
\begin{equation}
P_{0}=-\frac{r^{5}}{60G}
(4B_{00l}{}^{l}-S_{00l}{}^{l})=-\frac{r^{5}}{30G}(7E_{ab}E^{ab}-3H_{ab}H^{ab}).
\end{equation}
However, the Papapetrou pseudotensor gives a value which is a
linear combination of $B_{\alpha\beta\mu\nu}$ and
$V_{\alpha\beta\mu\nu}$ at the second order
\cite{SoCQG2009,SoNesterCQG2009}.  This indicates that there is a
chance to obtain a positive energy expression in the large region.
Recall the result for Papapetrou in vacuum:
\begin{equation}
2\kappa\,P^{\alpha\beta}=\frac{1}{9}(4B^{\alpha\beta}{}_{\mu\nu}-V^{\alpha\beta}{}_{\mu\nu})x^{\mu}x^{\nu}.
\end{equation}
Similarly, the corresponding gravitational energy from
(\ref{5Jan2009}) is
\begin{equation}
P_{0}=-\frac{r^{5}}{60G}
(4B_{00l}{}^{l}-V_{00l}{}^{l})=-\frac{r^{5}}{20G}(E_{ab}E^{ab}+H_{ab}H^{ab}).
\end{equation}

Before we go on to study any further, however, there comes a
question whether $V_{\alpha\beta\mu\nu}$ and
$B_{\alpha\beta\mu\nu}$ are totally equivalent?  Although they
have some components that are exactly the same (e.g.,
$V_{\mu{}000}\equiv{}B_{\mu{}000}$), we find that
$V_{\alpha\beta\mu\nu}$ and $B_{\alpha\beta\mu\nu}$ are indeed
different tensors \cite{SoCQG2009}. It is easy to clarify that
these two are different, since they are defined by different
fundamental quadratic curvatures, explicitly
\begin{eqnarray}
B_{\alpha\beta\mu\nu}&=&\tilde{B}_{\alpha\beta\mu\nu}+\tilde{T}_{\alpha\mu\nu},\\
V_{\alpha\beta\mu\nu}&=&\tilde{S}_{\alpha\beta\mu\nu}
+\tilde{K}_{\alpha\beta\mu\nu}+\tilde{T}_{\alpha\mu\nu}.\label{11Jan2009}
\end{eqnarray}
In particular, $V_{\alpha\beta\mu\nu}$ is completely trace free
but not completely symmetric \cite{SoCQG2009}.  The following
lists some properties
\begin{eqnarray}
&&V_{\alpha\beta\mu\nu}\equiv{}V_{(\alpha\beta)(\mu\nu)}\equiv{}V_{(\mu\nu)(\alpha\beta)},\quad{}
V_{\alpha\beta\mu}{}^{\mu}\equiv{}0\equiv{}V_{\alpha\mu\beta}{}^{\mu},\\
&&V_{0000}\equiv{}V_{00l}{}^{l}\equiv{}V_{m}{}^{m}{}_{l}{}^{l}\equiv{}V_{ml}{}^{ml}
\equiv{}E_{ab}E^{ab}+H_{ab}H^{ab}\equiv{}B_{0000},\\
&&V_{\mu{}000}\equiv{}V_{\mu{}0l}{}^{l}\equiv{}V_{\mu{}l0}{}^{l}
\equiv{}(E_{ab}E^{ab}+H_{ab}H^{ab},2\epsilon_{cab}E^{ad}H^{b}{}_{d})\equiv{}B_{\mu{}0l}{}^{l}.
\end{eqnarray}

It is known that $B_{\alpha\beta\mu\nu}$ has the dominant energy
property \cite{Penrose, Senovilla}
\begin{equation}
B_{\alpha\beta\mu\nu}\,w_{1}^{\alpha}\,w_{2}^{\beta}\,w_{3}^{\mu}\,w_{4}^{\nu}\geq{}0,
\label{13Jan2009}
\end{equation}
where $w_{1}, w_{2}, w_{3}, w_{4}$ are any future-pointing causal
vectors.   While $V_{\alpha\beta\mu\nu}$ only satisfies the weak
energy condition and $W_{\alpha\beta\mu\nu}$ fulfills none of
them.  However, we found $W_{\alpha\beta\mu\nu}$ has some
interesting properties. A simple computation using
(\ref{29April2009}) shows that
\begin{equation}
W_{\alpha\beta\mu\nu}u^{\alpha}t^{\beta}t^{\mu}t^{\nu}=0,\quad{}
W_{\alpha\beta\mu\nu}u^{\alpha}u^{\beta}u^{\mu}u^{\nu}=0,
\end{equation}
where $t$ is a timelike unit normal vector and $u$ can be timelike
or null.  Looking at (\ref{3aMar2010}), $V_{\alpha\beta\mu\nu}$
contains more information than $B_{\alpha\beta\mu\nu}$, however it
seems that $B_{\alpha\beta\mu\nu}$ is the important part of
$V_{\alpha\beta\mu\nu}$ and $W_{\alpha\beta\mu\nu}$ is a kind of
gauge freedom (i.e., it has no important physical effect).

A physical reasonable energy-momentum tensor should fulfill
certain energy conditions.  In particular, the local energy
density measured by the observer with a 4-velocity should be
non-negative. This energy condition must be true for all timelike
unit normal vectors \cite{Stephani}. In fact, we found
$V_{\alpha\beta\mu\nu}$ has the non-negative `energy' property
\begin{equation}
V_{\alpha\beta\mu\nu}t^{\alpha}t^{\beta}t^{\mu}t^{\nu}
\equiv{}B_{\alpha\beta\mu\nu}t^{\alpha}t^{\beta}t^{\mu}t^{\nu}
=E_{ab}E^{ab}+H_{ab}H^{ab}\geq{}0.
\end{equation}
Looking at this weak energy condition and from the reason of the
continuity \cite{Stephani}, the above inequalities must still be
true if the timelike vector $t$ is replaced by a null vector $v$.
Indeed, we found
\begin{equation}
V_{\alpha\beta\mu\nu}v^{\alpha}v^{\beta}v^{\mu}v^{\nu}
\equiv{}B_{\alpha\beta\mu\nu}v^{\alpha}v^{\beta}v^{\mu}v^{\nu}\geq{}0.\label{1Oct2009}
\end{equation}
Here we take a simple test, with 3 different cases, to evaluate
the inequality of (\ref{1Oct2009}) and get some idea what is the
value can be. Without loss of generality, consider 3 simple cases
from the unit normal null vector $v_{1}=k(1,1,0,0)$,
$v_{2}=k(1,0,1,0)$, $v_{3}=(1,0,0,1)$ and $k$ is a constant. We
found
\begin{equation}
B_{\alpha\beta\mu\nu}v^{\alpha}_{i}v^{\beta}_{i}v^{\mu}_{i}v^{\nu}_{i}
=k^{4}(B_{0000}+4B_{000i}+6B_{00ii}+4B_{0iii}+B_{iiii}),
\end{equation}
where $i=1,2,3$.  More precisely, we have calculated the results
referring to the five different distinct Petrov types
\cite{Petrov} as shown in Table {\ref{24Feb2010}}.

\begin{table}
\begin{tabular}{llll}
\hline
 & $v_{1}^{\alpha}=k(1,1,0,0)$ & $v_{2}^{\alpha}=k(1,0,1,0)$ & $v_{3}^{\alpha}= k(1,0,0,1)$\\
\hline {\mbox{Type I}} & $4k^{4}\left[
\begin{array}{cccc}
(E_{22}-E_{33})^{2}\\
+(H_{22}-H_{33})^{2}\\
\end{array}
\right]$ & $4k^{4}\left[
\begin{array}{cccc}
(E_{11}-E_{33})^{2}\\
+(H_{11}-H_{33})^{2}\\
\end{array}
\right]$ &
$4k^{4}\left[
\begin{array}{cccc}
(E_{11}-E_{22})^{2}\\
+(H_{11}-H_{22})^{2}\\
\end{array}
\right]$ \\
{\mbox{Type D}} & 0 & $9k^{4}(E_{11}^{2}+H_{11}^{2})$ &
$9k^{4}(E_{11}^{2}+H_{11}^{2})$ \\
 {\mbox{Type II}} & $64k^{4}(E_{23}^{2}+H_{23}^{2})$ &
$4k^{4}\left[
\begin{array}{cccc}
(E_{11}-E_{33})^{2}\\
+(H_{11}-H_{33})^{2}\\
\end{array}
\right]$ &
$4k^{4}\left[
\begin{array}{cccc}
(E_{11}-E_{22})^{2}\\
+(H_{11}-H_{22})^{2}\\
\end{array}
\right]$ \\
{\mbox{Type III}} & 0 & $16k^{4}(E_{12}^{2}+H_{12}^{2})$
& $16k^{4}(E_{12}^{2}+H_{12}^{2})$\\
{\mbox{Type N}} & $64k^{4}(E_{22}^{2}+H_{22}^{2})$ &
$4k^{4}(E_{22}^{2}+H_{22}^{2})$
& $4k^{4}(E_{22}^{2}+H_{22}^{2})$\\
\hline
\end{tabular}
\centering \caption{Five different Petrov types}\label{24Feb2010}
\end{table}

Note that different Petrov types corresponding to different values
simply because they are evaluated from different cannonial frames.
Likewise, it is not surprising that different frames associated
with different values from the same superpotential such as the
Freud superpotential \cite{SoIJMPD,Deser,Garecki,SoNesterCJP2009}
(i.e., using holonomic frames or orthonormal frames). The
statement is correct according to \cite{Stephani} for
$V_{\alpha\beta\mu\nu}$, which is based on the fact that
$B_{\alpha\beta\mu\nu}$ has the dominant energy property.

Following from (\ref{5Jan2009}), the energy-momentum density for
$V_{\alpha\beta\mu\nu}$ in the small sphere limit is
\begin{equation} 2\kappa\,P_{\mu}
=\frac{4\pi{}r^{5}}{15}\left(V^{0}{}_{\mu\alpha}{}^{\alpha}-V^{0}{}_{\mu0}{}^{0}\right)
=-\frac{4\pi{}r^{5}}{15}V_{0\mu{}00}.
\end{equation}
Or, more covariantly,
\begin{equation}
2\kappa\,P_{\mu}u^{\mu}
=-\frac{4\pi{}r^{5}}{15}\,V_{\mu\alpha\beta\gamma}u^{\mu}t^{\alpha}t^{\beta}t^{\gamma},
\end{equation}
where
\begin{equation}
V_{\alpha\beta\mu\nu}t^{\beta}t^{\mu}t^{\nu}\equiv{}
B_{\alpha\beta\mu\nu}t^{\beta}t^{\mu}t^{\nu}
=(E_{ab}E^{ab}+H_{ab}H^{ab},2\epsilon_{cab}E^{ad}H^{b}{}_{d}),
\end{equation}
and it should be recalled that
\begin{equation}
E_{ab}E^{ab}+H_{ab}H^{ab}\geq|2\epsilon_{cab}E^{ad}H^{b}{}_{d}|.
\end{equation}
The physical meaning (non-spacelike energy-momentum) is here
simpler and clearer than that of the dominant energy condition
(\ref{13Jan2009}). Obviously $V_{\alpha\beta\mu\nu}$ can play the
same role as $B_{\alpha\beta\mu\nu}$, it ensures a causal 4
momentum in the small sphere limit. In other words, the
`energy-momentum' density according to $B_{\alpha\beta\mu\nu}$ and
$V_{\alpha\beta\mu\nu}$ are on equal footing at the small sphere
region limit. Complementary, the Bel-Robinson tensor is no longer
the only tensor that have the unique preference for achieving the
causal 4 momentum in the quasilocal small region, but
$V_{\alpha\beta\mu\nu}$ can play the same role and it becomes the
unique alternative choice.

\subsection{Positive energy for the general fourth rank tensor $V'_{\alpha\beta\mu\nu}$}
From the technical point of view, if we are just interested in
positive energy and relax the restriction on the pseudotensor
constraint, which means the conservation of the energy-momentum,
there are an infinite number of combinations that have the weak
energy condition, not including $B_{\alpha\beta\mu\nu}$. We define
\begin{equation}
V'_{\alpha\beta\mu\nu}:=\tilde{K}_{\alpha\beta\mu\nu}+s\tilde{S}_{\alpha\beta\mu\nu}
+t_{1}\tilde{T}_{\alpha\beta\mu\nu}+t_{2}\tilde{T}_{\alpha\mu\beta\nu}
+t_{3}\tilde{T}_{\alpha\nu\beta\mu}.\label{9cSep2009}
\end{equation}
where $s, t_{1}, t_{2}, t_{3}$ are real numbers and
$t_{1}+t_{2}+t_{3}=1$.  Obviously, the energy-momentum
contribution for $\tilde{S}_{\alpha\beta\mu\nu}$ can be ignored
according to (\ref{29aApril2009}).  Explicitly
\begin{equation}
\tilde{S}_{\alpha\beta\mu\nu}u^{\alpha}t^{\beta}t^{\mu}t^{\nu}
\equiv0\equiv\tilde{S}_{\alpha\beta\mu\nu}u^{\alpha}u^{\beta}u^{\mu}u^{\nu}.
\end{equation}
On the other hand,
\begin{equation}
\tilde{T}_{\alpha\beta\mu\nu}u^{\alpha}u^{\beta}u^{\mu}u^{\nu}
\equiv\tilde{T}_{\alpha\mu\beta\nu}u^{\alpha}u^{\beta}u^{\mu}u^{\nu}
\equiv\tilde{T}_{\alpha\nu\beta\mu}u^{\alpha}u^{\beta}u^{\mu}u^{\nu}.
\end{equation}
Once again, here $u$ can be timelike or null.  Using
(\ref{11Jan2009}), rewrite (\ref{9cSep2009})
\begin{equation}
V'_{\alpha\beta\mu\nu}=V_{\alpha\beta\mu\nu}
+(s-1)\tilde{S}_{\alpha\beta\mu\nu}
+(t_{1}-1)\tilde{T}_{\alpha\beta\mu\nu}+t_{2}\tilde{T}_{\alpha\mu\beta\nu}
+t_{3}\tilde{T}_{\alpha\nu\beta\mu}.
\end{equation}
Based on the weak energy condition and continuity property, we
found
\begin{equation}
V'_{\alpha\beta\mu\nu}u^{\alpha}u^{\beta}u^{\mu}u^{\nu} \equiv{}
V_{\alpha\beta\mu\nu}u^{\alpha}u^{\beta}u^{\mu}u^{\nu} \equiv{}
B_{\alpha\beta\mu\nu}u^{\alpha}u^{\beta}u^{\mu}u^{\nu}\geq{}0.
\end{equation}
This illustrates that there does exist an infinite number of
combinations which have positivity if we exclude the conservation
of the energy-momentum requirement according to the pseudotensor
restriction.

Furthermore, in order to obtain the dominant energy condition, the
Bel-Robinson tensor is the unique tensor that has the suitable
combination from the four fundamental quadratic curvature
combinations, namely from (\ref{20aJan2009}) to
(\ref{20bJan2009}). As a matter of fact, $B_{\alpha\beta\mu\nu}$
has more nice properties than the other quadratic curvature
combinations generally (e.g., $S_{\alpha\beta\mu\nu}$ and
$K_{\alpha\beta\mu\nu}$). In particular, $B_{\alpha\beta\mu\nu}$
possesses the completely symmetric property. However, concerning
the gravitational energy at the small sphere limit, we found that
$V_{\alpha\beta\mu\nu}$ is the unique alternative choice to
compare with $B_{\alpha\beta\mu\nu}$.

\section{Conclusion}
Using the four fundamental quadratic curvature tensors, we
constructed all the possible combinations in the quasilocal small
sphere region expression. We recovered that
$B_{\alpha\beta\mu\nu}$ gives a definite positive gravitational
energy (more previously a causal 4-momentum) in the small sphere
limit approximation. However, we found an unique alternative, the
recently proposed tensor $V_{\alpha\beta\mu\nu}$, which also
contributes the same non-negative gravitational energy density at
the same region limit.  Based on the two physical conditions:
energy-momentum conservation and casuality.  We found that these
two tensors can be classified as a basis for expressions which
have the desirable non-negative gravitational energy in the small
sphere region.  In other words, $B_{\alpha\beta\mu\nu}$ and
$V_{\alpha\beta\mu\nu}$ are on equal footing in the small sphere
limit.  This means that if we obtain $B_{\alpha\beta\mu\nu}$ or
$V_{\alpha\beta\mu\nu}$ from the gravitational expression at the
small scale, either of them is good enough to search whether the
expression is positive or not at the large scale.  For example,
the Papapetrou pseudotensor can be a good candidate to study the
positivity energy expression, as it is proportional to the linear
combination of $B_{\alpha\beta\mu\nu}$ and $V_{\alpha\beta\mu\nu}$
at the second order evaluation.

We found that only a linear combination of $B_{\alpha\beta\mu\nu}$
and $V_{\alpha\beta\mu\nu}$ satisfies the energy-momentum
conservation and casuality physical conditions.  Remarkably, the
completely trace free property for both $B_{\alpha\beta\mu\nu}$
and $V_{\alpha\beta\mu\nu}$ demonstrates the same two physical
requirements.  It may be interesting and also from the practical
reason to count the independent components for
$B_{\alpha\beta\mu\nu}$, $V_{\alpha\beta\mu\nu}$ and
$W_{\alpha\beta\mu\nu}$. We found that there are 25 independent
components for $B_{\alpha\beta\mu\nu}$, 35 components for
$V_{\alpha\beta\mu\nu}$ and 10 for $W_{\alpha\beta\mu\nu}$.
Moreover, we found that $W_{\alpha\beta\mu\nu}$, associated with
$V_{\alpha\beta\mu\nu}$, behaves as a kind of gauge freedom.
Furthermore, relaxing the restriction of the energy-momentum
conservation requirement for the pseudotensor,
$V'_{\alpha\beta\mu\nu}$ demonstrates that there are an infinite
number of ways to obtain positivity, namely the weak energy
condition. For the conserved expressions $B_{\alpha\beta\mu\nu}$
satisfies the dominant energy condition while
$V_{\alpha\beta\mu\nu}$ does not, but does fulfill the weak energy
condition.

\section*{Acknowledgment}
This work was supported by NSC 95-2811-M-032-008, NSC
96-2811-M-032-001, NSC 97-2811-M-032-007  and NSC
98-2811-M-008-078.

\end{document}